\documentclass[10pt,conference]{IEEEtran}
\usepackage{amsmath,amsfonts}
\usepackage{graphicx,algorithm,algpseudocode}
\usepackage{cite}
\usepackage{xcolor}
\usepackage[font=small]{caption}
\usepackage{subcaption}
\usepackage{amsfonts,booktabs}
\usepackage{multirow,verbatim}
\usepackage{color, colortbl}
\usepackage{lipsum}
\usepackage{soul}
\usepackage[normalem]{ulem}

\newcolumntype{L}[1]{>{\raggedright\let\newline\\\arraybackslash\hspace{0pt}}m{#1}}
\newcolumntype{C}[1]{>{\centering\let\newline\\\arraybackslash\hspace{0pt}}m{#1}}
\newcolumntype{R}[1]{>{\raggedleft\let\newline\\\arraybackslash\hspace{0pt}}m{#1}}

\IEEEoverridecommandlockouts
\allowdisplaybreaks

\begin{document}
\bstctlcite{IEEEexample:BSTcontrol}

\title{I-SCOUT: Integrated Sensing and Communications to Uncover Moving Targets in NextG Networks}

\author{
\IEEEauthorblockN{Utku Demir\IEEEauthorrefmark{1}, Kemal Davaslioglu\IEEEauthorrefmark{1}, Yalin E. Sagduyu\IEEEauthorrefmark{1}, Tugba Erpek\IEEEauthorrefmark{1}, Gustave Anderson\IEEEauthorrefmark{2}, Sastry Kompella\IEEEauthorrefmark{1}} \\ \IEEEauthorrefmark{1}Nexcepta Inc., Gaithersburg, MD, USA \\
\IEEEauthorrefmark{2}Lockheed Martin, Bethesda, MD, USA
}

\maketitle
\thispagestyle{plain}
\pagestyle{plain}

\begin{abstract}
Integrated Sensing and Communication (ISAC) represents a transformative approach within 5G and beyond, aiming to merge wireless communication and sensing functionalities into a unified network infrastructure. This integration offers enhanced spectrum efficiency, real-time situational awareness, cost and energy reductions, and improved operational performance. ISAC provides simultaneous communication and sensing capabilities, enhancing the ability to detect, track, and respond to spectrum dynamics and potential threats in complex environments. In this paper, we introduce I-SCOUT, an innovative ISAC solution designed to uncover moving targets in NextG networks. We specifically repurpose the Positioning Reference Signal (PRS) of the 5G waveform, exploiting its distinctive autocorrelation characteristics for environment sensing. The reflected signals from moving targets are processed to estimate both the range and velocity of these targets using the cross ambiguity function (CAF). We conduct an in-depth analysis of the tradeoff between sensing and communication functionalities, focusing on the allocation of PRSs for ISAC purposes. Our study reveals that the number of PRSs dedicated to ISAC has a significant impact on the system's performance, necessitating a careful balance to optimize both sensing accuracy and communication efficiency. Our results demonstrate that I-SCOUT effectively leverages ISAC to accurately determine the range and velocity of moving targets. Moreover, I-SCOUT is capable of distinguishing between multiple targets within a group, showcasing its potential for complex scenarios. These findings underscore the viability of ISAC in enhancing the capabilities of NextG networks, for both commercial and tactical applications where precision and reliability are critical.
\end{abstract}


\section{Introduction}
\label{sec:introduction}
The evolution of 5G technology and its progression towards future generations mark significant advancements in connectivity, with the potential to revolutionize both commercial and tactical operations through the convergence of communication and sensing capabilities. Integrated Sensing and Communication (ISAC) aims to leverage the same network infrastructure for both data transmission and environmental sensing, enhancing spectrum efficiency and operational effectiveness \cite{xu2022edge,demirhan2022integrated,
zhangSurvey,xiong2023, liu2022integrated, cui2021integrating,liu2022survey,wild2021joint, wei2023integrated}. By utilizing incumbent communication signals for sensing tasks, ISAC can improve both the resolution and accuracy of situational awareness, while optimizing communication performance through enhanced environmental understanding.

ISAC is poised to revolutionize a broad spectrum of applications across both commercial and defense sectors. In the commercial domain, ISAC can enhance smart city initiatives by enabling advanced traffic management systems, automated driving, and improved public safety through pervasive environmental monitoring. The ability to simultaneously communicate and sense the environment opens new possibilities for industrial automation, where accurate target detection and real-time data transmission are crucial.

In defense applications, ISAC is indispensable for significantly enhancing situational awareness and operational effectiveness. The ability to detect, track, and classify moving targets in real-time provides a tactical advantage, crucial for modern warfare and defense strategies. ISAC within a single framework reduces the need for separate systems, streamlining operations and reducing the logistical burden on tactical forces. Integrating ISAC into defense networks offers shared infrastructure, leading to cost and energy savings, and the potential for high-resolution sensing using advanced communication frequencies. These benefits are particularly relevant in defense scenarios, where real-time data exchange and accurate sensing are essential for mission success.

\begin{figure}[t!]
\centering
  \includegraphics[width=0.85\linewidth]{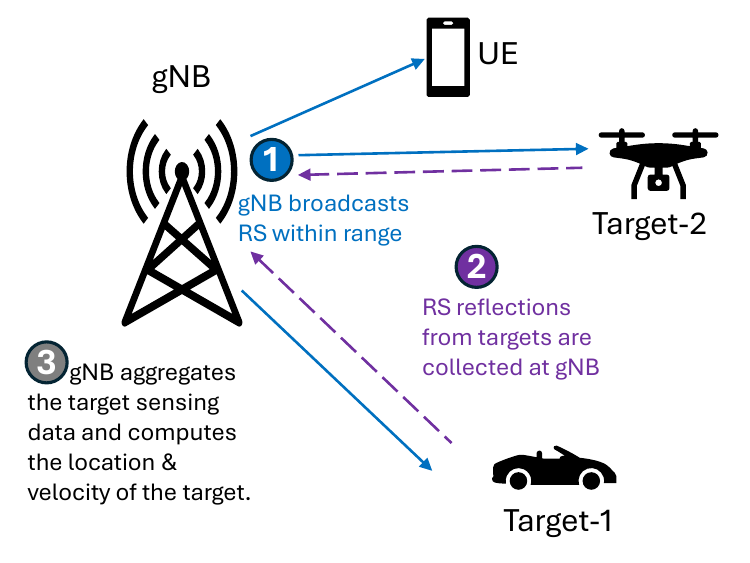}
    \caption {The workflow of the I-SCOUT framework for radar application within 5G NR network summarized in three steps.}
    \vspace{-2mm}
  \label{fig:Fig1}
\end{figure}

\begin{figure*}
\centering
  \includegraphics[width=0.85\linewidth]{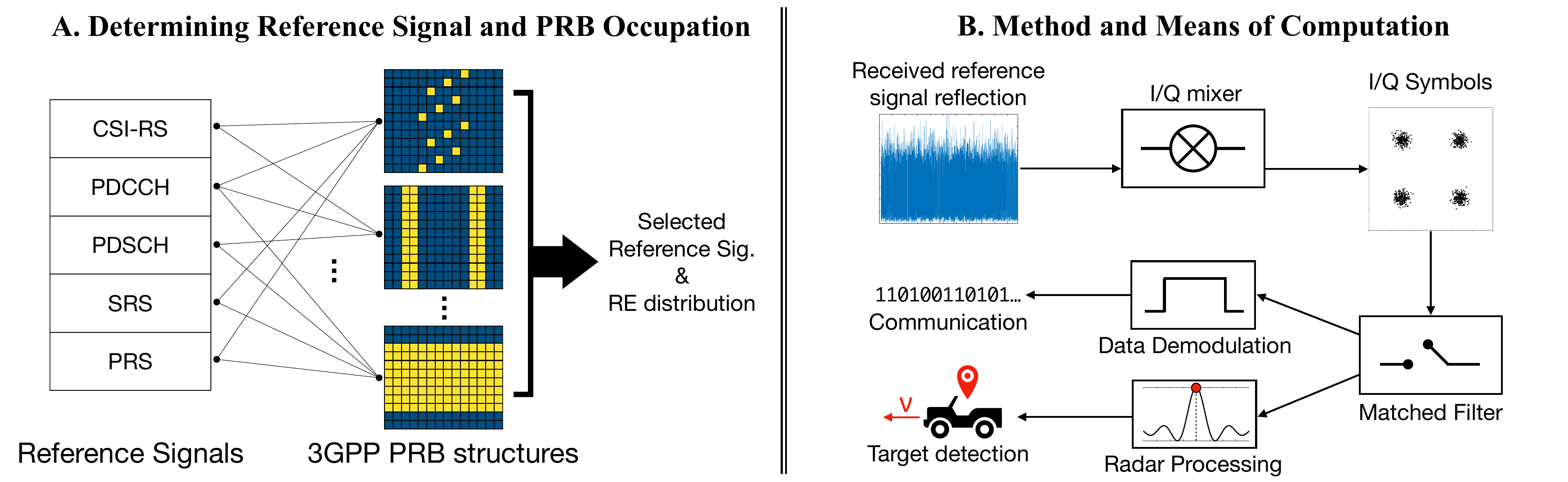}
    \caption {Important aspects to realize ISAC capabilities in 5G and beyond networks.}
    \vspace{-2mm}
  \label{fig:Fig2}
\end{figure*}

In this paper, we propose a framework to utilize the already available 5G network for ISAC without putting extra burden on the cellular network, which we achieve by using the reference signals (RS) transmitted from the base station (gNB) to user equipments (UEs). As shown in Fig.~\ref{fig:Fig1}, we introduce I-SCOUT: Integrated Sensing and Communications to Uncover Moving Targets in NextG Networks. For ISAC, it is important to determine reference signal and PRB occupation (Fig.~\ref{fig:Fig2}.A) and determine the method and mean of computation for sensing (Fig.~\ref{fig:Fig2}.B). I-SCOUT leverages the Positioning Reference Signal (PRS) of the 5G waveform, known for its distinctive autocorrelation characteristics, to perform environmental sensing. The technical process involves transmitting PRS signals, which, upon reflection from moving targets, are captured and processed to estimate the range and velocity of these targets using the cross ambiguity function (CAF) to enable precise target detection and tracking. Reflected signals can be captured at multiple locations, such as the original or another base station, or the UE. However, involving another BS would require coordination, while using the UE would introduce additional computational demands and necessitate hardware and software modifications to access the reflected data. Therefore, we choose to use the original BS for capturing the reflected signals..

One of the core benefits of I-SCOUT is its ability to balance the tradeoff between sensing and communication functionalities. By carefully selecting the number of PRSs allocated for ISAC, I-SCOUT ensures high sensing accuracy, reflected in low Root Mean Squared Error (RMSE), without compromising communication efficiency. This balance is critical for applications where both high-quality communication and accurate sensing are required simultaneously.

The novel contributions of this paper are multifold:
\begin{enumerate}
\item Repurposing PRS for Sensing: This work demonstrates a novel application of the 5G PRS for environment sensing, showcasing its potential beyond traditional communication roles.
\item Integration with CAF: The use of CAF for processing reflected PRS signals represents an advanced approach to estimate range and velocity with high precision.
\item Tradeoff Analysis: A detailed study on the tradeoff between sensing and communication functionalities provides insights into optimal resource allocation, enhancing the practical applicability of ISAC systems.
\item Experimental Validation: We demonstrate the viability of I-SCOUT approach and highlight its capability to accurately estimate the range and velocity of targets and distinguish among multiple targets.
\end{enumerate}

The remainder of the paper is organized as follows, Sec.~\ref{sec:RelatedWork} discusses related work. Sec.~\ref{sec:SystemModel} describes the system model including 5G New Radio (NR) reference signals, 5G NR waveform, and sensing resource allocation. Sec.~\ref{sec:ProposedSolution} introduces the range and velocity estimation solution.  Sec~\ref{sec:Results} presents the performance evaluation. Sec.~\ref{sec:Conclusion} concludes the paper.

\section{Related Work}
\label{sec:RelatedWork}
5G networks can be leveraged for environment sensing using monostatic and bistatic radar techniques depending on whether transmitter and receiver of signals are co-located or separated, respectively. 5G network was utilized in \cite{samczynski20215g} to estimate range and velocity of a target in a bistatic radar scenario, where a new type of waveform is utilized for channel illumination that is based on Synchronization Signal Block's (SSB) Primary Synchronization Signal (PSS) in a USRP testbed. The target velocity and range were estimated using the compliant PRS in a monostatic radar scenario in~\cite{wei20225g}. In comparison, I-SCOUT offers dynamic adjustment of PRS resource blocks in a monostatic setting to balance the sensing-communication performance tradeoffs. The extension to multistatic scenario was considered in \cite{li2023towards}, where the probability of detection of a sensing coverage region is maximized with the use of multiple UEs acting as receiver ends.

One emerging application of ISAC is in vehicle-to-everything (V2X) networks due to spectrum scarcity and sensing criticality. A reference signal was designed in~\cite{zhao2023reference} for V2X applications by considering reference signal spacing, power allocation, and subcarrier spacing in an optimization framework. The impact of radio access communication parameters, such as bandwidth, modulation and coding scheme, packet size, on the sensing performance was studied in~\cite{10438868} in terms of detection capability and parameter estimation accuracy when multiple vehicle interferers are present. The performance on range and velocity estimation with and without inter-vehicle interference was studied in \cite{bartoletti2022sidelink}.

To support emerging 6G applications, ISAC was also considered jointly with task-oriented and semantic communications in a multi-task learning framework, optimizing target sensing and data transfer while preserving semantic information \cite{sagduyu2024will,sagduyu2023joint}. Sensing component of ISAC can be also combined with other modalities such as computer vision to improve reliability of target sensing \cite{sagduyu2023multijoint, lu2024semantic}.

\section{System Model}
\label{sec:SystemModel}

\subsection{5G NR Reference Signals}
The 5G NR physical channel includes several reference signals: Positioning Reference Signal (PRS), Demodulation Reference Signal (DM-RS), Channel-State Information Reference Signal (CSI-RS), and Synchronization Signal (SS) \cite{3gpp.38.211,3gpp.38.214}. PRS is used for accurate, real-time node location \cite{3gpp.38.214}. DM-RS aids in decoding the Physical Downlink Shared Channel (PDSCH). CSI-RS is transmitted by the gNB for the UE to estimate downlink channel quality and manage beams. SS is used in the 5G NR synchronization procedure for the UE to detect and decode primary/secondary synchronization signals (PSS/SSS) and the physical broadcast channel (PBCH). PRS, CSI-RS, DMRS, and SSS use Gold sequences for their strong correlation, enabling signal demultiplexing without synchronization. In this paper, we use PRS as its resource block allocation can be dynamically adjusted.

\subsection{5G NR Waveform}
The downlink, uplink, and sidelink transmissions are organized into subframes each consisting of 1~msec duration. Ten subframes make up a frame of 10~msec duration. Transmission parameters of the 5G NR waveform are flexible and can be adjusted depending on the need. The 5G numerology that is used to define the subcarrier spacing, the OFDM symbol, cyclic prefix (CP), and the total duration of the OFDM symbol including the CP is shown in Table~\ref{table:5g_numerology}. In this table, $\Delta_f$ represents the subcarrier spacing, while the OFDM symbol duration, cyclic prefix (CP) duration, and OFDM symbol including CP are denoted by $T$, $T_{CP}$, and $T_s = T + T_{CP}$, respectively. As the numerology increases, the OFDM symbol and the CP durations decrease, resulting in a shorter overall OFDM symbol duration including CP. 

\begin{table}[tb!]
\centering
\caption{5G NR numerology for transmission parameters \cite{3gpp.38.214}.}
\label{table:5g_numerology}
\begin{tabular}{c|c|c|c|c}
\toprule
$\mu$ & $\Delta_f$ & $T$ ($\mu$sec) & $T_{CP}$ ($\mu$sec) & $T_s$ ($u$sec) \\ \midrule
   0 & 15 & 66.67 & 4.69 & 71.35 \\ 
   1 & 30 & 33.33 & 2.34 & 35.68 \\ 
   2 & 60 & 16.67 & 1.17 & 17.84 \\ 
   3 & 120 & 8.33 & 0.57 & 8.92 \\ 
   4 & 240 & 4.17 & 0.29 & 4.46 \\ 
   \bottomrule
\end{tabular}
\end{table}

Using the 5G NR waveform, we use a  resource grid of $N_{\textit{PRB}} N_{\textit{SC}}^{\textit{PRB}}$ subcarriers and $N_{symb}$ OFDM symbols are defined, where $N_{\textit{\textit{PRB}}}$ represents the number of allocated PRBs, $N_{\textit{SC}}^{\textit{PRB}}$ denotes the number of subcarriers per resource block, and $N_{symb}$ stands for the number of OFDM symbols per subframe. Using this notation, as an example, we can express the PRS signal as a continuous time-domain signal which is defined as 
\begin{align}
    x_T(t) = \sum_{m=0}^{M-1} \sum_{k=0}^{N_j-1} s(k,m) \exp\left(j 2\pi f_k t\right) u\bigg(\frac{t-mT_s}{T_s}\bigg),
\end{align}
where $s(k,m)$ represents the modulated PRS symbol with subcarrier $k$ and OFDM symbol index $m$. The number of OFDM symbols in the time domain is expressed as $M$ and $N_j$ is the number of subcarriers carrying the PRS signals. The frequency for the $k$th subcarrier is represented by $f_k$, and $u(t/T_s)$ denotes the rectangular function. 

The resource elements (RE) allocated to PRS are equally spaced in frequency domain such that $f_k$ satisfies \cite{3gpp.38.214}
\begin{align}
    f_k = (K_{\textit{comb}}^{\textit{PRS}} \times k + k_0) \Delta_f, \quad k=0,\ldots,N_j-1,
\end{align}
where $K_{\textit{comb}}^{\textit{PRS}}$ is the comb size of PRS, which defines the RE density of all PRS resources in a PRS resource set and it can take the values of $K_{\textit{comb}}^{\textit{PRS}} \in \{2,4,6,12\}$. This means that every $K_{\textit{comb}}^{\textit{PRS}}$th RE in the PRB is allocated for the PRS. The starting index of the first PRS subcarrier is represented by $k_0$. 

\subsection{Sensing Resource Allocation}
The gNB needs to dynamically allocate the frequency and time resources to provide services such as synchronization, channel estimation, and downlink data payload for communication. To sense the surrounding environment, the gNB needs to allocate some of its resources for sensing. In this paper, we propose to use the reference signals that are defined in the 5G NR waveform (e.g., 3GPP 38.211 \cite{3gpp.38.211}) and process these signal returns to sense the targets in the environment. For example, in the 5G NR, the PRS signal can be used to estimate the range and velocities of the targets within the coverage area of a set of gNBs. In this paper, we explore the opportunity to use these signals to provide sensing capability for any target (not only the UEs) that is mobile within the coverage area. This application has several important use cases such as beam prediction in the millimeter wave, enhanced perimeter security and surveillance, smart traffic management, industrial automation, emergency response, enhanced retail experiences, and smart city applications. Depending on the need (e.g., number of targets to track and their physical distribution), the gNB can adjust the number of PRBs allocated for sensing.

We consider a gNB that sends transmissions for communication and sensing services. The transmitted waveform propagates and the reflections from the targets are  received at the gNB. The return signal (echo) is processed at the gNB, which can be expressed as 
\begin{align}
    x_R(t,m)  =  & \hspace{0.5em} \xi \cdot x_T(t,m) \exp\left(-j 2\pi K_{\textit{comb}}^{\textit{PRS}} k \Delta_f (2 R_l/c)\right) \nonumber \\ & \times
    \exp\left(j 2\pi K_{\textit{comb}}^{\textit{PRS}} mT_s f_{d,l}\right),
\end{align}
where $\xi$ is the attenuation factor, $R_l$ and $f_{d,l}$ are the range and  Doppler shift of target $l$, respectively. The speed of light is denoted by $c$. Note that since the signal travels from the gNB to target and back to the gNB, we consider the two-way propagation and we can express the Doppler shift as 
\begin{align}
    f_{d,r} = \frac{2 v_{rel,l}}{\lambda} = \frac{2 v_{l} f_c}{c},
\end{align}
where $\lambda$ is the wavelength and $v_{rel,l}$ is the relative velocity of target $l$. Since the gNB is considered to be stationary, we drop the relative and simply denote it as $v_l$.

\section{Range and Velocity Estimation in I-SCOUT}
\label{sec:ProposedSolution}
To remove the transmitted signal information from the received signal, we convolve $x_R(t,m)$ with the reference signal and improve the signal-to-noise ratio (SNR). We then apply Fast Fourier Transform (FFT) along the fast-time dimension (range dimension) to transform the time-domain signal into the frequency domain. This helps in separating the signals reflected from different ranges. Similarly, we apply FFT along the slow-time dimension (Doppler dimension) to transform the signal into the Doppler domain to estimate the relative velocity of targets. The outputs of the range FFT and Doppler FFT form a 2D matrix representing the Range-Doppler map. 

To detect targets, we apply a threshold to the range-Doppler map to detect peaks, which indicate the presence of targets. Using the threshold, we identify the locations of the peaks in the Range-Doppler map. The coordinates of these peaks correspond to the range and Doppler shift of the targets. As the final step, we obtain the range and velocity of the targets from the peak locations. We express the maximum unambiguous range, $R_{\max}$, in terms of 5G waveform parameters as 
\begin{align}
R_{\max} = \frac{c N_j}{2N \Delta_f} = \frac{c}{2 K_{\textit{comb}}^{\textit{PRS}} \Delta_f},    
\end{align}
and the range resolution $\Delta_R$ can be expressed as 
\begin{align}
    \Delta_R = \frac{c}{2N_{\textit{PRB}} \Delta_f}.
    \label{eqn:resol_R}
\end{align}

\begin{table}[t!]
    \centering
    \caption{Parameter definitions and values}
    \label{tab:notation}
    \small
    {\begin{tabular}{c|c|c}
    \hline
        Notation & Description & Values \\
        \hline
        $f_c$ & Carrier frequency & $24$,$2.5$~GHz  \\
        $B$ & Signal bandwidth & $\{10,...,100\}$~MHz  \\
        $\Delta_f$ & Subcarrier spacing & $\{30, 60, 120, 240\}$~kHz \\
        $P_{\textit{Tx}}$ & gNB transmit power & $10^{-2}$~W  \\
        $c$ & Speed of light & $3\times10^{8}\,\mathrm{m/s}$  \\
        $M$ & Number of transmitted  &  128 \\
        & OFDM symbols &   \\
        \hline
    \end{tabular}}    
\end{table}

For the velocity estimation, the maximum unambiguous velocity can be defined as 
\begin{align}
    \nu_{\max} = \frac{c M_j}{2 M T_s f_c} = \frac{c}{2 K_{\textit{comb}}^{\textit{PRS}}T_s f_c},
\end{align}
and the velocity resolution is 
\begin{align}
    \Delta_{\nu} = \frac{c}{2 M T_s f_c},
    \label{eqn:resol_v}
\end{align} 
where $M$ is the number of transmitted OFDM symbols. The intuition of our dynamic PRS waveform adaptation approach is when the targets are closely separated but still can be resolved separately, that is, they are separated larger than $\Delta R$, they will appear as one in the Range-Doppler if we use a small $N_{\textit{PRB}}$. In that case, we can increase the resources allocated to sensing and improve the sensing performance. If the number of targets still remains the same as we allocate more PRBs for PRS, then we can stop. For this purpose, we employ a binary search algorithm to determine the number of PRBs to use for PRS. We initialize the search boundaries $n_L=0$ and $n_R=62$ and set the current value as $n_M=31$. The number of PRBs is obtained $N_{\textit{PRB}} = 4 n_M+24$. This ensures that we do not violate the conditions in \cite{3gpp.38.211}. We measure the number of targets. In the second step, we increase $n_M$ to $\lfloor (n_M+n_H)/2 \rfloor =46$ and $N_{\textit{PRB}}=208$ and re-count the number of targets. If the target count has increased, we set $n_L=n_M$ and if the target count remains the same, we set $n_R=n_M$. Finally, we update $n_M = \lfloor(n_M+n_L)/2\rfloor$ and repeat.

For performance evaluation, we measure the RMSE of range estimations, $e_R$, which can be expressed as 
\begin{align}
    e_{R} = \Bigg( \frac{1}{N_{\textit{\textit{target}}}} \sum_{l=1}^{N_{\textit{\textit{target}}}} \left(r_{true}(l) - \hat{r}(l)\right)^2 \Bigg)^{1/2},
    \label{eqn:rmse_r}
\end{align}
where $r_{true}(l)$ and $\hat{r}(l)$ stand for the true and estimated range of target $l$, respectively. Similarly, we can represent the  RMSE for velocity estimation, $e_{\nu}$, as 
\begin{align}
    e_{\nu} = \Bigg( \frac{1}{N_{\textit{\textit{target}}}} \sum_{l=1}^{N_{\textit{\textit{target}}}} \left(\nu_{true}(l) - \hat{\nu}(l)\right)^2 \Bigg)^{1/2},
    \label{eqn:rmse_v}
\end{align}
where $v_{true}(l)$ and $\hat{v}(l)$ represent for the true and estimated range of target $l$, respectively. Using (\ref{eqn:rmse_r}) and~(\ref{eqn:rmse_v}), we define normalized sensing error as
\begin{align}
    \bar{e} = \frac{1}{2} \left(\frac{e_{\nu}}{\max(e_{\nu})} + \frac{e_{R}}{\max(e_{R})}\right).
    \label{eqn:normSensError}
\end{align}

\begin{figure}[t!]
\centering
  \includegraphics[width=0.85\linewidth]{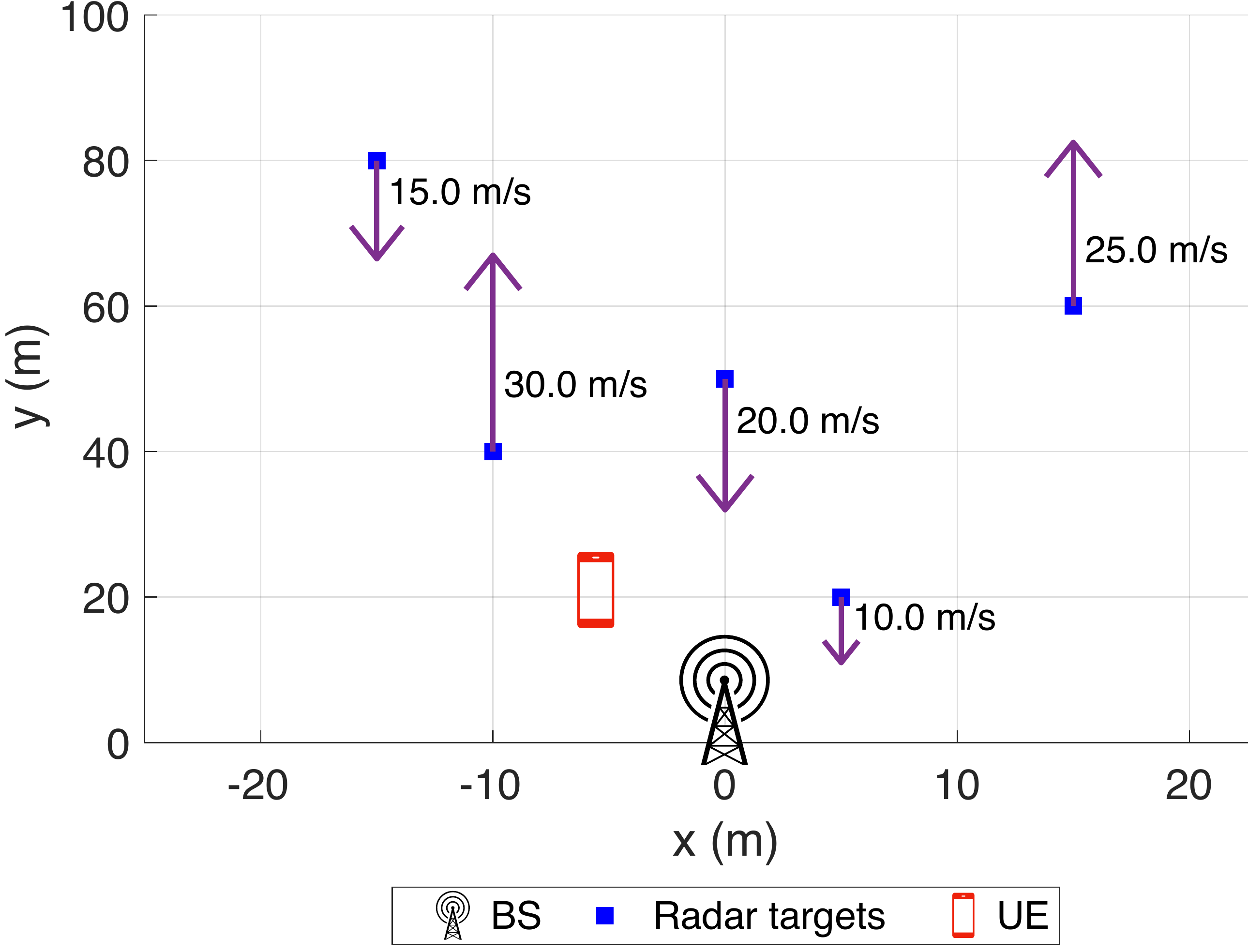}
    \caption {An example scenario, displaying 5 targets, their trajectories, a UE, and a gNB with sensing capabilities.}
  \label{fig:Fig3}
\end{figure}

\begin{figure*}[t!]
     \centering
     \begin{subfigure}[b]{0.3\textwidth}
         \centering
         \includegraphics[width=\textwidth]{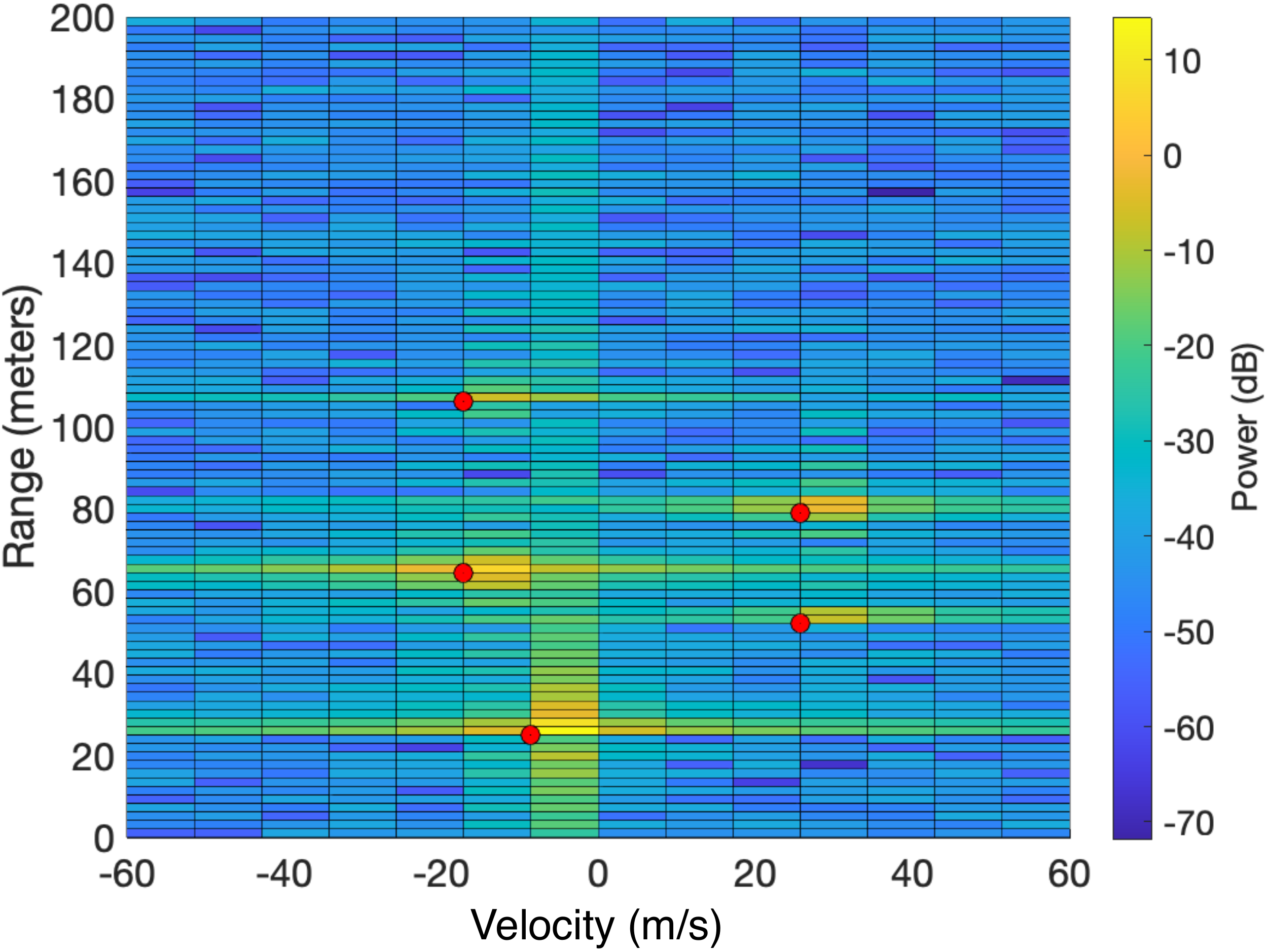}
         \caption{$N_{\textit{PRB}} = 36$ for $\Delta f = 240$kHz.}
         \label{fig:Fig4a}
     \end{subfigure}
     \hfill
     \begin{subfigure}[b]{0.3\textwidth}
         \centering
         \includegraphics[width=\textwidth]{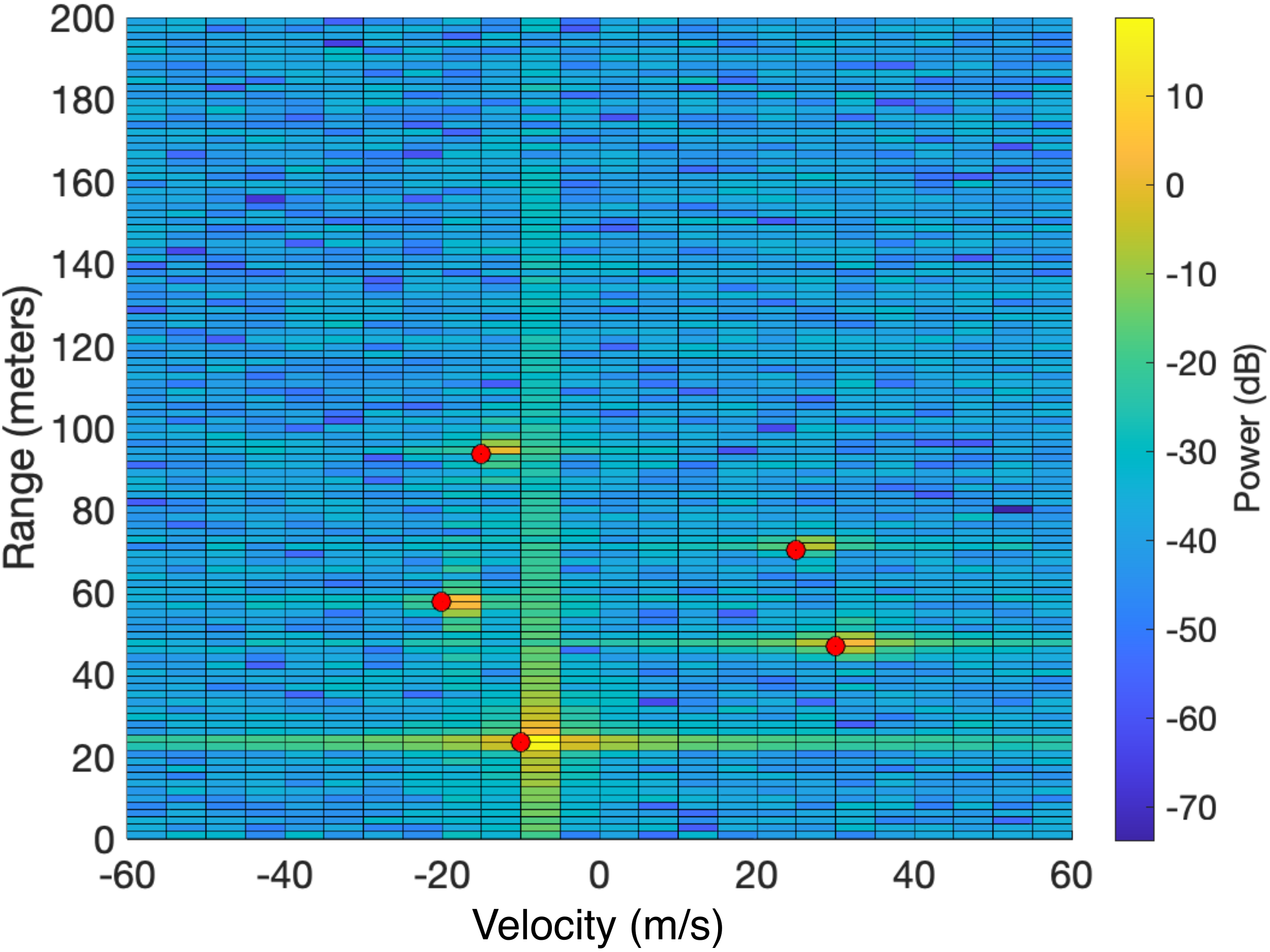}
         \caption{$N_{\textit{PRB}} = 68$ for $\Delta f = 120$kHz.}
         \label{fig:Fig4b}
     \end{subfigure}
     \hfill
     \begin{subfigure}[b]{0.3\textwidth}
         \centering
         \includegraphics[width=\textwidth]{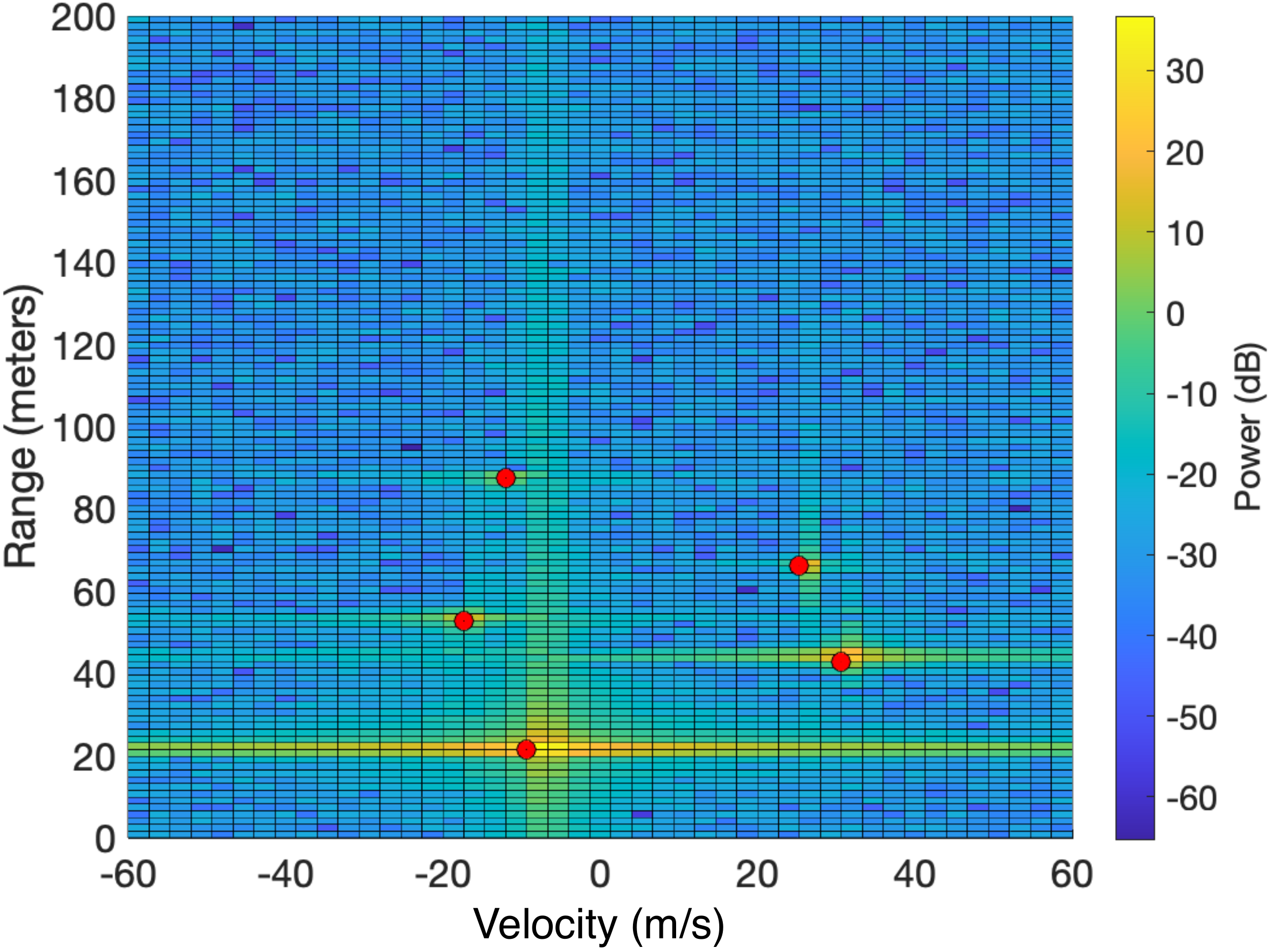}
         \caption{$N_{\textit{PRB}} = 140$ for $\Delta f = 60$kHz.}
         \label{fig:Fig4c}
     \end{subfigure}
        \caption{Range-Doppler spectrums with $B = 100$~MHz and $f_c = 24$~GHz (FR2 band) as $\Delta_f$ varies.}
        \label{fig:Fig4}
\end{figure*}

\begin{figure*}
     \centering
     \begin{subfigure}[b]{0.24\textwidth}
         \centering
         \includegraphics[width=\textwidth]{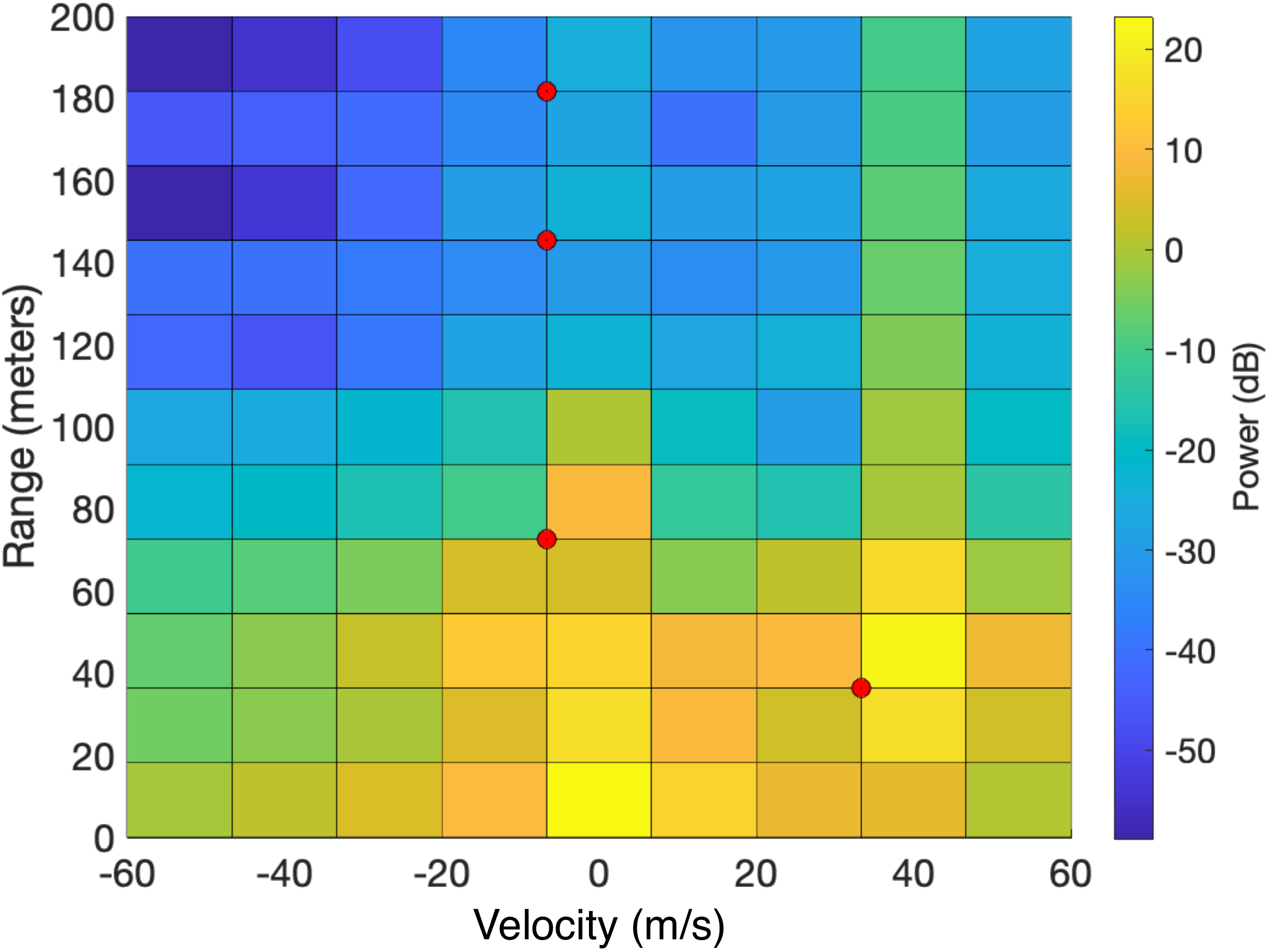}
         \caption{$N_{\textit{PRB}} = 28$ for $B = 10$~MHz.}
         \label{fig:Fig5a}
     \end{subfigure}
     \hfill
     \begin{subfigure}[b]{0.24\textwidth}
         \centering
         \includegraphics[width=\textwidth]{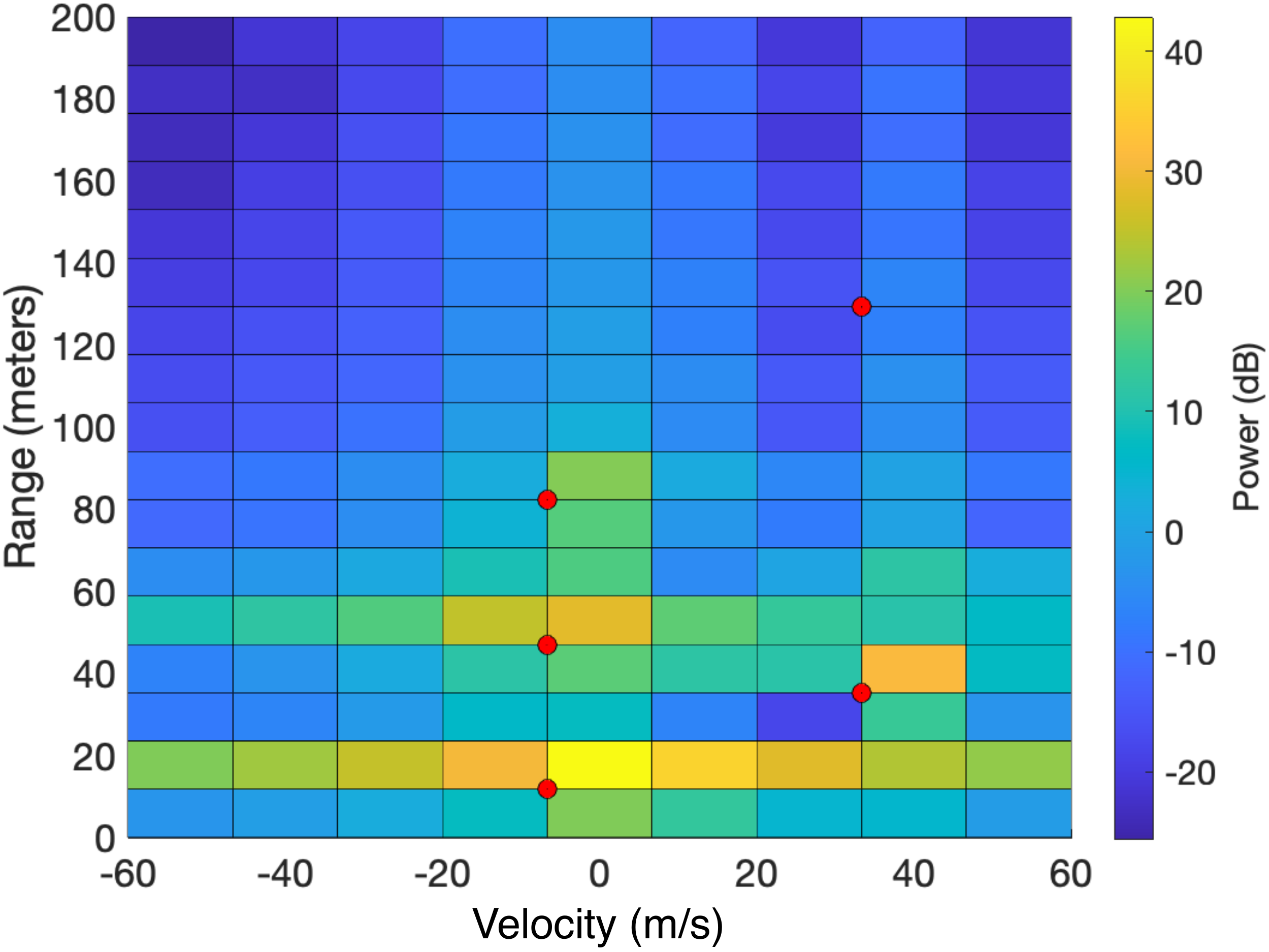}
         \caption{$N_{\textit{PRB}} = 40$ for $B = 15$~MHz.}
         \label{fig:Fig5b}
     \end{subfigure}
     \hfill
     \begin{subfigure}[b]{0.24\textwidth}
         \centering
         \includegraphics[width=\textwidth]{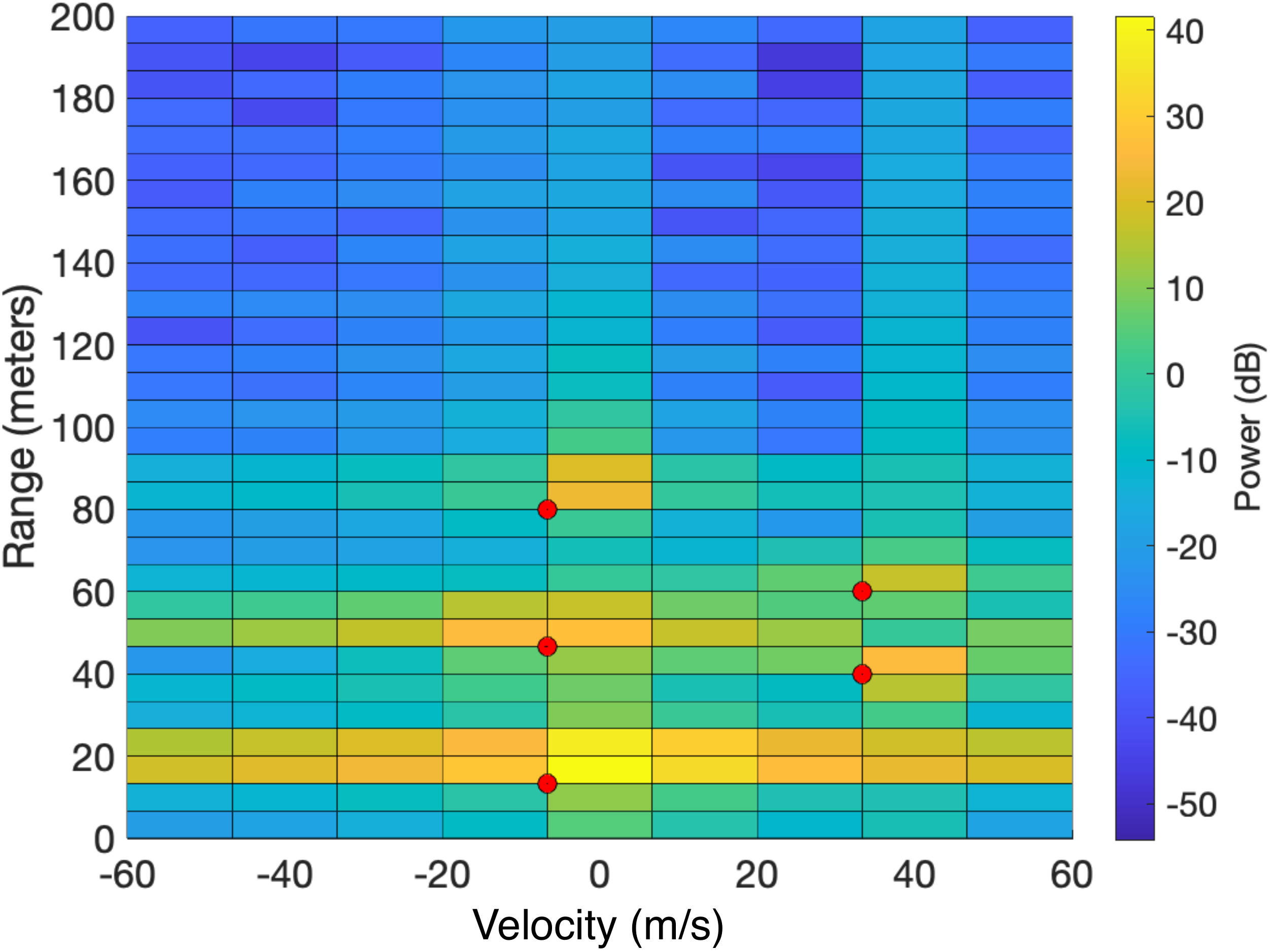}
         \caption{$N_{\textit{PRB}} = 68$ for $B = 25$~MHz.}
         \label{fig:Fig5c}
     \end{subfigure}
     \hfill
     \begin{subfigure}[b]{0.24\textwidth}
         \centering
         \includegraphics[width=\textwidth]{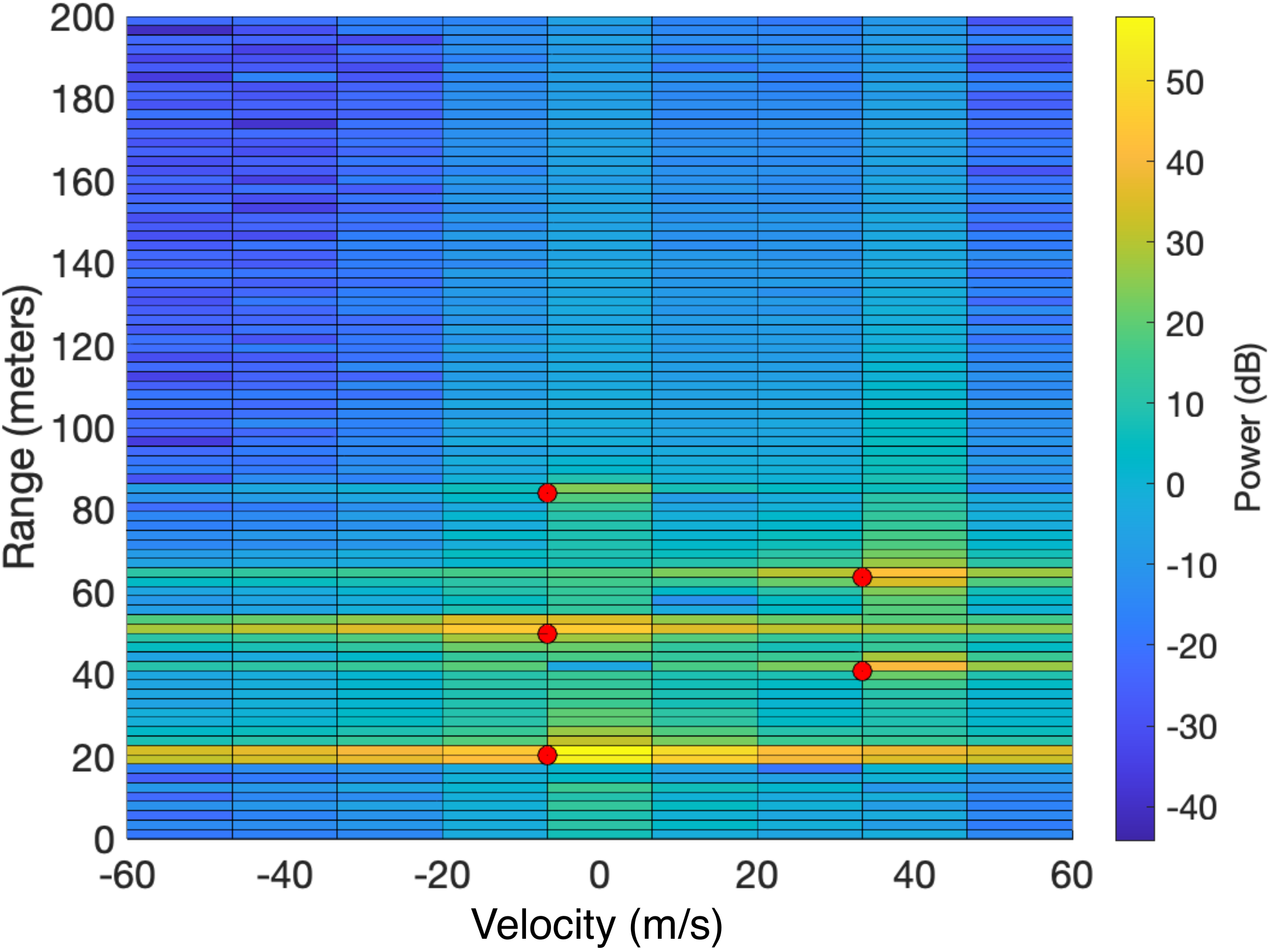}
         \caption{$N_{\textit{PRB}} = 224$ for $B = 80$~MHz.}
         \label{fig:Fig5d}
     \end{subfigure}
        \caption{Range-Doppler spectrums with $\Delta_f = 30$~kHz and $f_c = 2.5$~GHz (FR1 band) as $B$ varies.}
        \label{fig:Fig5}
\end{figure*}

\section{Performance Evaluation}
\label{sec:Results}
\subsection{Simulation Setup}
The performance evaluation of I-SCOUT involves multiple radar targets at various ranges and set velocities. I-SCOUT can be applied to capture an arbitrary number of targets without any prior information. In this paper, we evaluate the performance in a scenario given in Fig.~\ref{fig:Fig3}, where there are five targets (e.g., drones), one gNB, and one UE to represent a simplified 5G network deployment. For the simplicity, we let the targets follow a trajectory parallel to the y-axis. Moving targets are modeled with the Radar Cross Section (RCS) of 4~$\mathrm{m}^2$. The waveform, channel, and experiment parameters are provided in Table~\ref{tab:notation}. We set up two experiments, one in the Frequency Range 1 (FR1) and another one in the Range 2 (FR2) of the 5G~NR standard. For the FR1 experiments, we use the carrier frequency of $f_c=2.5$~GHz and bandwidth of $B=\{10,15,20,..., 90\}$~MHz). For the FR2 experiments, we use the carrier frequency of $24$~GHz and bandwidth of $100$~MHz.

As indicated in Table~\ref{table:5g_numerology}, $\Delta_f$ can take values of $15\times2^k$~kHz, where $k\in\{0,1,2,3,4\}$. Also, the number of PRB for reference signals could be in between 24 and 272 with 4 increments. Given that $\Delta_f = B / (N_{\textit{PRB}}\times12)$, $12$ being the number of subcarriers per PRB, in order to meet the aforementioned constraints, we need to choose $N_{\textit{PRB}}\in\{36, 68, 140\}$ in the FR1 experiments to meet $\Delta_f\in\{60,120,240\}$ kHz. $\Delta_f$ values of $15$~kHz and $30$~kHz in this configuration would not comply with the 5G standard as the required $N_{\textit{PRB}}$ would exceed 272. In fact, only three choices of $\Delta_f$ are possible for a given $B$ in FR2. With this configuration, 100 simulations are conducted with five targets by assigning different trajectories to each of them, where the maximum assigned velocity and range values are $20$~m/s and $102$~m, respectively. For the FR1 experiments, $\Delta_f=30$~kHz is used.

\subsection{Velocity and Range Estimation} 
Velocity and range estimation is performed through examining Range-Doppler spectrum for which we provide a series of examples in Figs.~\ref{fig:Fig4} and~\ref{fig:Fig5} for the FR2 and FR1 experiments, respectively. The red dots indicate the detected targets, which lead to the estimation of the targets' velocity and range. In the FR2 experiment results, shown in Fig.~\ref{fig:Fig4}, as we decrease the subcarrier spacing, with the constant $B$, velocity resolution decreases (more sample points within the same interval), making its estimation more robust against estimation errors. The red dots on the Range-Doppler spectrum display the algorithmically detected local maxima points. By visual inspection, we observe that the located peaks are very close to the true ranges and velocities, i.e., the distance between the detected target and the gNB, given in Fig.~\ref{fig:Fig3}, across all the possible 5G compliant OFDM waveforms, considering the selected $f_c$ and $B$.

We display a few examples of the 5G NR FR1 experiments in Fig.~\ref{fig:Fig5}. Note that as the increase the $B$ and keep $\Delta_f$ constant, the range resolution decreases (more sample points within the same interval), making its estimation more robust against estimation errors. As in the FR2 experiments, detected peaks are very close to the true velocities and ranges.

\begin{figure}[tbh!]
\vspace{-0.25cm}
\centering
  \includegraphics[width=0.63\linewidth]{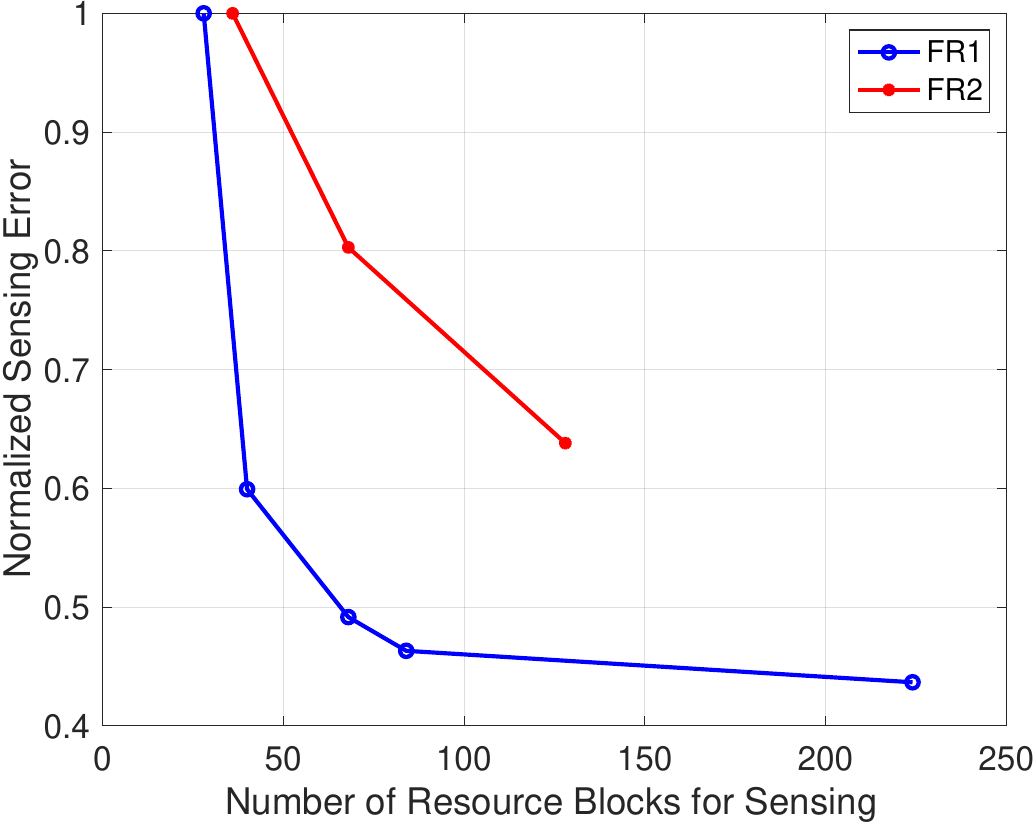}
    \caption {Sensing error vs. number of resource block for sensing.}
    \vspace{-2mm}
  \label{fig:Fig6}
\end{figure}

\begin{figure}[tbh!]
\centering
  \includegraphics[width=0.63\linewidth]{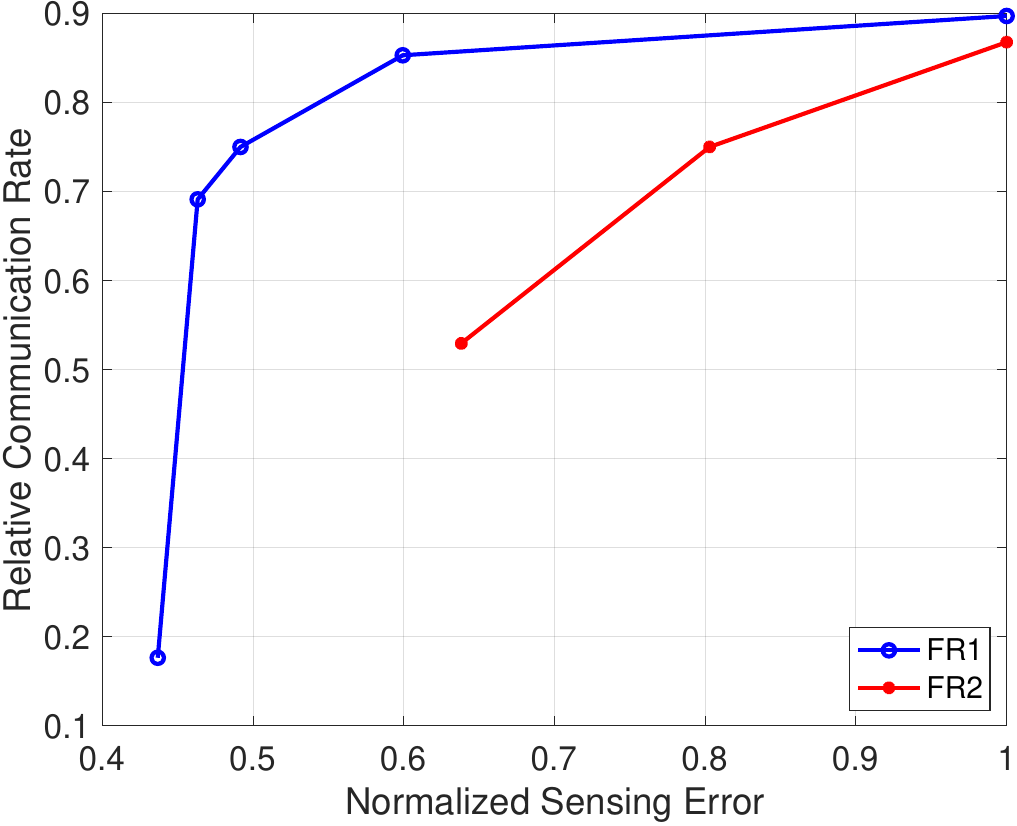}
    \caption {Communication rate vs. sensing error.}
    \vspace{-2mm}
  \label{fig:Fig7}
\end{figure}

We quantitatively demonstrate the range and velocity estimation performances in Fig.~\ref{fig:Fig6} and how this affects the communication within a network in Fig.~\ref{fig:Fig7}. The respective markers in these figures are read from Fig.~\ref{fig:Fig4} and~\ref{fig:Fig5}. Note that the normalized sensing errors (Eq.~\ref{eqn:normSensError}) in both FR1 and FR2 experiments decrease as we increase $N_{\textit{PRB}}$, as expected.

In Fig.~\ref{fig:Fig7}, we show how the selected $N_{\textit{PRB}}$ affects the relative communication rate, as unused PRB could be used for communication purposes, i.e., $(N_{\textit{PRB}}^{\max}-N_{\textit{PRB}})/N_{\textit{PRB}}^{\max}$, where $N_{\textit{PRB}}^{\max}=272$. We notice that in both FR1 and FR2 experiments communication rate increases with the normalized sensing error. The reason is that less $N_{\textit{PRB}}$ causes more error, but in turn, it leaves more PRBs available for communication purposes. Considering the target detection accuracy and communication demand in a 5G network, an optimal value for $N_{RB}$ could be determined as discussed in Sec.~\ref{sec:ProposedSolution}. The common axis between Fig.~\ref{fig:Fig6} and~\ref{fig:Fig7} is the normalized sensing error. From Fig.~\ref{fig:Fig6}, we observe that for the same normalized error, more RBs for sensing are needed for FR2, thereby reducing the communication opportunity.

\section{Conclusion}
\label{sec:Conclusion}
In this paper, we introduce I-SCOUT, an innovative solution designed to integrate sensing and communication functionalities within NextG networks, leveraging the capabilities of 5G and beyond communication systems. I-SCOUT is based on repurposing the PRSs of the 5G waveform for environment sensing, exploiting its distinctive autocorrelation characteristics. I-SCOUT processes the reflected PRS signals from moving targets to estimate both their range and velocity using the CAF. We conduct a comprehensive analysis of the tradeoff between sensing and communication functionalities, particularly in terms of the number of PRSs utilized for sensing purposes in ISAC. We demonstrate the effectiveness of I-SCOUT in accurately determining the range and velocity of moving targets and distinguishing between multiple targets within a group. In addition, our results show how the use of PRSs jointly affect the sensing and communication performance, leading to a design space to utilize for NextG networks. 

\bibliographystyle{IEEEtran}
\bibliography{ISCOUT_arXiv_References}

\end{document}